\documentclass{article}

\PassOptionsToPackage{numbers, compress, sort}{natbib}

\usepackage[final]{nips_2017}
\usepackage[utf8]{inputenc} 
\usepackage[T1]{fontenc}    
\usepackage{hyperref}       
\usepackage{url}            
\usepackage{booktabs}       
\usepackage{amsfonts}       
\usepackage{amsmath}       	
\usepackage{nicefrac}       
\usepackage{microtype}      
\usepackage[capitalise]{cleveref}
\usepackage{graphicx}
\usepackage{soul}
\usepackage{xcolor}

\usepackage[font={footnotesize}]{caption}

\usepackage{macros}

\title{Beyond Parity: \\Fairness Objectives for Collaborative Filtering}

\author{
  Sirui Yao\\
  Department of Computer Science\\
 Virginia Tech\\
  Blacksburg, VA 24061 \\
  \texttt{ysirui@vt.edu} \\
\And
  Bert Huang\\
  Department of Computer Science\\
 Virginia Tech\\
  Blacksburg, VA 24061 \\
  \texttt{bhuang@vt.edu} \\
}

\begin{document}

\maketitle

\begin{abstract}
  We study fairness in collaborative-filtering recommender systems, which are sensitive to discrimination that exists in historical data. Biased data can lead collaborative-filtering methods to make unfair predictions for users from minority groups. We identify the insufficiency of existing fairness metrics and propose four new metrics that address different forms of unfairness. These fairness metrics can be optimized by adding fairness terms to the learning objective. Experiments on synthetic and real data show that our new metrics can better measure fairness than the baseline, and that the fairness objectives effectively help reduce unfairness.
\end{abstract}

\section{Introduction}

This paper introduces new measures of unfairness in algorithmic recommendation and demonstrates how to optimize these metrics to reduce different forms of unfairness.
Recommender systems study user behavior and make recommendations to support decision making. They have been widely applied in various fields to recommend items such as movies, products, jobs, and courses.  However, since recommender systems make predictions based on observed data, they can easily inherit bias that may already exist. To address this issue, we first formalize the problem of unfairness in recommender systems and identify the insufficiency of demographic parity for this setting. We then propose four new unfairness metrics that address different forms of unfairness. We compare our fairness measures with non-parity on biased, synthetic training data and prove that our metrics can better measure unfairness. To improve model fairness, we provide five fairness objectives that can be optimized, each adding unfairness penalties as regularizers. Experimenting on real and synthetic data, we demonstrate that each fairness metric can be optimized without much degradation in prediction accuracy, but that trade-offs exist among the different forms of unfairness.

We focus on a frequently practiced approach for recommendation called collaborative filtering, which makes recommendations based on the ratings or behavior of other users in the system. The fundamental assumption behind collaborative filtering is that other users' opinions can be selected and aggregated in such a way as to provide a reasonable prediction of the active user's preference \citep{ekstrand2011collaborative}. For example, if a user likes item A, and many other users who like item A also like item B, then it is reasonable to expect that the user will also like item B. Collaborative filtering methods would predict that the user will give item B a high rating.

With this approach, predictions are made based on co-occurrence statistics, and most methods assume that the missing ratings are missing at random. Unfortunately, researchers have shown that sampled ratings have markedly different properties from the users' true preferences \citep{marlin2012collaborative, marlin2009collaborative}. Sampling is heavily influenced by social bias, which results in more missing ratings in some cases than others. This non-random pattern of missing and observed rating data is a potential source of unfairness. For the purpose of improving recommendation accuracy, there are collaborative filtering models \citep{marlin2012collaborative, beutel2017beyond, sahebi2015takes}  that use side information to address the problem of imbalanced data, but in this work, to test the properties and effectiveness of our metrics, we focus on the basic matrix-factorization algorithm first. Investigating how these other models could reduce unfairness is one direction for future research.

Throughout the paper, we consider a running example of unfair recommendation. We consider recommendation in education, and unfairness that may occur in areas with current gender imbalance, such as science, technology, engineering, and mathematics (STEM) topics. Due to societal and cultural influences, fewer female students currently choose careers in STEM. For example, in 2010, women accounted for only 18\% of the bachelor's degrees awarded in computer science \citep{broad2014recruiting}. The underrepresentation of women causes historical rating data of computer-science courses to be dominated by men. Consequently, the learned model may underestimate women's preferences and be biased toward men. We consider the setting in which, even if the ratings provided by students accurately reflect their true preferences, the bias in which ratings are reported leads to unfairness.

The remainder of the paper is organized as follows. First, we review previous relevant work in \cref{sec:related}. In \cref{sec:approach}, we formalize the recommendation problem, and we introduce four new unfairness metrics and give justifications and examples. In \cref{sec:experiments}, we show that unfairness occurs as data gets more imbalanced, and we present results that successfully minimize each form of unfairness. Finally, \cref{sec:conclusion} concludes the paper and proposes possible future work.

\section{Related Work}
\label{sec:related}

As machine learning is being more widely applied in modern society, researchers have begun identifying the criticality of algorithmic fairness. Various studies have considered algorithmic fairness in problems such as supervised classification \citep{pedreshi2008discrimination,lum2016statistical,zafar2017fairness}. When aiming to protect algorithms from treating people differently for prejudicial reasons, removing sensitive features (e.g., gender, race, or age) can help alleviate unfairness but is often insufficient. Features are often correlated, so other unprotected attributes can be related to the sensitive features and therefore still cause the model to be biased \citep{kamishima2011fairness, zemel2013learning}. Moreover, in problems such as collaborative filtering, algorithms do not directly consider measured features and instead infer latent user attributes from their behavior.

Another frequently practiced strategy for encouraging fairness is to enforce \emph{demographic parity}, which is to achieve statistical parity among groups. The goal is to ensure that the overall proportion of members in the protected group receiving positive (or negative) classifications is identical to the proportion of the population as a whole \citep{zemel2013learning}. For example, in the case of a binary decision $\hat{Y} \in \{0, 1\}$ and a binary protected attribute $A \in \{0, 1\}$, this constraint can be formalized as \citep{hardt2016equality}
\begin{equation}
\Pr\{\hat{Y} =1 | A = 0\} = \Pr\{\hat{Y} =1 | A = 1\}~.
\label{eq:parity}
\end{equation}

Kamishima et al.~\citep{kamishima2011fairness,kamishima2012enhancement,kamishima2013efficiency,kamishima2014correcting,kamishima2016model} evaluate model fairness based on this non-parity unfairness concept, or try to solve the unfairness issue in recommender systems by adding a regularization term that enforces demographic parity. The objective penalizes the differences among the average predicted ratings of user groups. However, demographic parity is only appropriate when preferences are unrelated to the sensitive features. In tasks such as recommendation, user preferences are indeed influenced by sensitive features such as gender, race, and age \citep{chausson2010watches, daymont1984job}. Therefore, enforcing demographic parity may significantly damage the quality of recommendations.

To address the issue of demographic parity, Hardt et al.~\citep{hardt2016equality} propose to measure unfairness with the true positive rate and true negative rate. This idea encourages what they refer to as \emph{equal opportunity} and no longer relies on the implicit assumption of demographic parity that the target variable is independent of sensitive features. They propose that, in a binary setting, given a decision $\hat{Y} \in \{0, 1\}$, a protected attribute $A \in \{0, 1\}$, and the true label  $Y \in \{0, 1\}$, the constraints are equivalent to \citep{hardt2016equality}
\begin{equation}
\Pr\{\hat{Y} =1 | A = 0, Y = y\} = \Pr\{\hat{Y} =1 | A = 1, Y = y\}, y  \in  \{0, 1\}~.
\label{eq:equalopportunity}
\end{equation}
This constraint upholds fairness and simultaneously respects group differences. It penalizes models that only perform well on the majority groups. This idea is also the basis of the unfairness metrics we propose for recommendation.

Our running example of recommendation in education is inspired by the recent interest in using algorithms in this domain \citep{sacin2009recommendation, thai2010recommender, dascalu2016educational}. Student decisions about which courses to study can have significant impacts on their lives, so the usage of algorithmic recommendation in this setting has consequences that will affect society for generations. Coupling the importance of this application with the issue of gender imbalance in STEM \citep{beede2011women} and challenges in retention of students with backgrounds underrepresented in STEM \citep{smith2011women,griffith2010persistence}, we find this setting a serious motivation to advance scientific understanding of unfairness---and methods to reduce unfairness---in recommendation.

\section{Fairness Objectives for Collaborative Filtering}
\label{sec:approach}

This section introduces fairness objectives for collaborative filtering. We begin by reviewing the matrix factorization method. We then describe the various fairness objectives we consider, providing formal definitions and discussion of their motivations.

\subsection{Matrix Factorization for Recommendation}

We consider the task of collaborative filtering using matrix factorization \citep{koren2009matrix}. We have a set of users indexed from 1 to $\numusers$ and a set of items indexed from 1 to $\numitems$. For the $i$th user, let $\group_i$ be a variable indicating which group the $i$th user belongs to. For example, it may indicate whether user $i$ identifies as a woman, a man, or with a non-binary gender identity. For the $j$th item, let $\itemgroup_j$ indicate the item group that it belongs to. For example, $\itemgroup_j$ may represent a genre of a movie or topic of a course. Let $\rating_{ij}$ be the preference score of the $i$th user for the $j$th item. The ratings can be viewed as entries in a rating matrix $\ratingmat$.

The matrix-factorization formulation builds on the assumption that each rating can be represented as the product of vectors representing the user and item. With additional bias terms for users and items, this assumption can be summarized as follows:
\begin{equation}
\rating_{ij} \approx \uservec_i ^\top \itemvec_j + \userbias_i + \itembias_j ~ ,
\label{eq:reconstruction}
\end{equation}
where $\uservec_i$ is a $d$-dimensional vector representing the $i$th user, $\itemvec_j$ is a $d$-dimensional vector representing the $j$th item, and $\userbias_i$ and $\itembias_j$ are scalar bias terms for the user and item, respectively. The matrix-factorization learning algorithm seeks to learn these parameters from observed ratings $\trainingdata$, typically by minimizing a regularized, squared reconstruction error:
\begin{equation}
J(\usermat, \itemmat, \userbiasvec, \itembiasvec) = \frac{\lambda}{2} \left( ||\usermat||^2_{\mathrm{F}} + ||\itemmat||^2_{\mathrm{F}} \right) + \frac{1}{|\trainingdata|} \sum_{(i, j) \in \trainingdata} \left( \prediction_{ij} - \rating_{ij} \right)^2 ~,
\label{eq:MF-objective}
\end{equation}
where $\userbiasvec$ and $\itembiasvec$ are the vectors of bias terms, $|| \cdot ||_{\mathrm{F}}$ represents the Frobenius norm, and
\begin{equation}
\prediction_{ij} = \uservec_i ^\top \itemvec_j + \userbias_i + \itembias_j.
\end{equation}
Strategies for minimizing this non-convex objective are well studied, and a general approach is to compute the gradient and use a gradient-based optimizer. In our experiments, we use the Adam algorithm \citep{kingma2014adam}, which combines adaptive learning rates with momentum.

\subsection{Unfair Recommendations from Underrepresentation}
\label{sec:example}

In this section, we describe a process through which matrix factorization leads to unfair recommendations, even when rating data accurately reflects users' true preferences. Such unfairness can occur with imbalanced data. We identify two forms of underrepresentation: \emph{population imbalance} and \emph{observation bias}. We later demonstrate that either leads to unfair recommendation, and both forms together lead to worse unfairness. In our discussion, we use a running example of course recommendation, highlighting effects of underrepresentation in STEM education.

Population imbalance occurs when different types of users occur in the dataset with varied frequencies. For example, we consider four types of users defined by two aspects. First, each individual identifies with a gender. For simplicity, we only consider binary gender identities, though in this example, it would also be appropriate to consider men as one gender group and women and all non-binary gender identities as the second group. Second, each individual is either someone who enjoys and would excel in STEM topics or someone who does and would not. Population imbalance occurs in STEM education when, because of systemic bias or other societal problems, there may be significantly fewer women who succeed in STEM (WS) than those who do not (W), and because of converse societal unfairness, there may be more men who succeed in STEM (MS) than those who do not (M). This four-way separation of user groups is not available to the recommender system, which instead may only know the gender group of each user, but not their proclivity for STEM. 

Observation bias is a related but distinct form of data imbalance, in which certain types of users may have different tendencies to rate different types of items. This bias is often part of a feedback loop involving existing methods of recommendation, whether by algorithms or by humans. If an individual is never recommended a particular item, they will likely never provide rating data for that item. Therefore, algorithms will never be able to directly learn about this preference relationship. In the education example, if women are rarely recommended to take STEM courses, there may be significantly less training data about women in STEM courses.

We simulate these two types of data bias with two stochastic block models \citep{holland1976local}. We create one block model that determines the probability that an individual in a particular user group likes an item in a particular item group. The group ratios may be non-uniform, leading to population imbalance. We then use a second block model to determine the probability that an individual in a user group rates an item in an item group. Non-uniformity in the second block model will lead to observation bias.

Formally, let matrix $\rateblockmat \in [0,1]^{|\group| \times |\itemgroup|}$ be the block-model parameters for rating probability. For the $i$th user and the $j$th item, the probability of $\rating_{ij}=+1$ is $\rateblock_{(\group_i, \itemgroup_j)}$, and otherwise $\rating_{ij} = -1$. Morever, let $\takeblockmat \in [0,1]^{|\group| \times |\itemgroup|}$ be such that the probability of observing $\rating_{ij}$ is $\takeblock_{(\group_i, \itemgroup_j)}$.

\subsection{Fairness Metrics}

In this section, we present four new unfairness metrics for preference prediction, all measuring a discrepancy between the prediction behavior for disadvantaged users and advantaged users. Each metric captures a different type of unfairness that may have different consequences. We describe the mathematical formulation of each metric, its justification, and examples of consequences the metric may indicate. We consider a binary group feature and refer to disadvantaged and advantaged groups, which may represent women and men in our education example. 

The first metric is \emph{value unfairness}, which measures inconsistency in signed estimation error across the user types, computed as
\begin{equation}
\metric_\val = \frac{1}{\numitems} \sum_{j = 1}^\numitems \left| \left( \avgpredF_j - \avgrateF_j \right) - \left( \avgpredM_j - \avgrateM_j \right) \right| ~ ,
\label{eq:value-unfairness}
\end{equation}
where $\avgpredF_j$ is the average predicted score for the $j$th item from disadvantaged users, $\avgpredM_j$ is the average predicted score for advantaged users, and $\avgrateF_j$ and $\avgrateM_j$ are the average ratings for the disadvantaged and advantaged users, respectively. Precisely, the quantity $\avgpredF_j$ is computed as
\begin{equation}
\avgpredF_j := \frac{1}{|\{ i: \left( (i, j) \in \trainingdata \right) \wedge  \group_i \}|} \sum_{i: \left( (i, j) \in \trainingdata \right) \wedge  \group_i } \prediction_{ij} ~,
\end{equation}
and the other averages are computed analogously.

Value unfairness occurs when one class of user is consistently given higher or lower predictions than their true preferences. If the errors in prediction are evenly balanced between overestimation and underestimation or if both classes of users have the same direction and magnitude of error, the value unfairness becomes small. Value unfairness becomes large when predictions for one class are consistently overestimated and predictions for the other class are consistently underestimated. For example, in a course recommender, value unfairness may manifest in male students being recommended STEM courses even when they are not interested in STEM topics and female students not being recommended STEM courses even if they are interested in STEM topics.

The second metric is \emph{absolute unfairness}, which measures inconsistency in absolute estimation error across user types, computed as
\begin{equation}
\metric_\absolute = \frac{1}{\numitems} \sum_{j = 1}^\numitems \left| \left| \avgpredF_j - \avgrateF_j \right| - \left| \avgpredM_j - \avgrateM_j \right| \right| ~.
\label{eq:abs-unfairness}
\end{equation}
Absolute unfairness is unsigned, so it captures a single statistic representing the quality of prediction for each user type. If one user type has small reconstruction error and the other user type has large reconstruction error, one type of user has the unfair advantage of good recommendation, while the other user type has poor recommendation. In contrast to value unfairness, absolute unfairness does not consider the direction of error. For example, if female students are given predictions 0.5 points below their true preferences and male students are given predictions 0.5 points above their true preferences, there is no absolute unfairness. Conversely, if female students are given ratings that are off by 2 points in either direction while male students are rated within 1 point of their true preferences, absolute unfairness is high, while value unfairness may be low.

The third metric is \emph{underestimation unfairness}, which measures inconsistency in how much the predictions underestimate the true ratings:
\begin{equation}
\metric_\underest = \frac{1}{\numitems} \sum_{j = 1}^\numitems \left| \max \{ 0, \avgrateF_j - \avgpredF_j \} - \max \{ 0, \avgrateM_j - \avgpredM_j \} \right| ~.
\label{eq:under-unfairness}
\end{equation}
Underestimation unfairness is important in settings where missing recommendations are more critical than extra recommendations. For example, underestimation could lead to a top student not being recommended to explore a topic they would excel in.

Conversely, the fourth new metric is \emph{overestimation unfairness}, which measures inconsistency in how much the predictions overestimate the true ratings:
\begin{equation}
\metric_\overest = \frac{1}{\numitems} \sum_{j = 1}^\numitems \left| \max \{ 0, \avgpredF_j - \avgrateF_j \} - \max \{ 0, \avgpredM_j - \avgrateM_j \} \right| ~.
\label{eq:over-unfairness}
\end{equation}
Overestimation unfairness may be important in settings where users may be overwhelmed by recommendations, so providing too many recommendations would be especially detrimental. For example, if users must invest large amounts of time to evaluate each recommended item, overestimating essentially costs the user time. Thus, uneven amounts of overestimation could cost one type of user more time than the other.

Finally, a \emph{non-parity} unfairness measure based on the regularization term introduced by Kamishima et al.~\citep{kamishima2011fairness} can be computed as the absolute difference between the overall average ratings of disadvantaged users and those of advantaged users:
\[
\metric_\parity = \left| \avgpredF - \avgpredM \right|~.
\]

Each of these metrics has a straightforward subgradient and can be optimized by various subgradient optimization techniques. We augment the learning objective by adding a smoothed variation of a fairness metric based on the Huber loss \citep{huber1964robust}, where the outer absolute value is replaced with the squared difference if it is less than 1. We solve for a local minimum, i.e,
\begin{equation}
\min_{\usermat, \itemmat, \userbiasvec, \itembiasvec}
 ~ J(\usermat, \itemmat, \userbiasvec, \itembiasvec) +
\metric ~.
\label{eq:fullobj}
\end{equation}
The smoothed penalty helps reduce discontinuities in the objective, making optimization more efficient. It is also straightforward to add a scalar trade-off term to weight the fairness against the loss. In our experiments, we use equal weighting, so we omit the term from \cref{eq:fullobj}.

\section{Experiments}
\label{sec:experiments}

We run experiments on synthetic data based on the simulated course-recommendation scenario and real movie rating data \citep{harper2016movielens}. For each experiment, we investigate whether the learning objectives augmented with unfairness penalties successfully reduce unfairness.

\subsection{Synthetic Data}

In our synthetic experiments, we generate simulated course-recommendation data from a block model as described in \cref{sec:example}. We consider four user groups $\group \in \{\textrm{W}, \textrm{WS}, \textrm{M}, \textrm{MS}\}$ and three item groups $\itemgroup \in \{ \textrm{Fem}, \textrm{STEM}, \textrm{Masc} \}$. The user groups can be thought of as women who do not enjoy STEM topics (W), women who do enjoy STEM topics (WS), men who do not enjoy STEM topics (M), and men who do (MS). The item groups can be thought of as courses that tend to appeal to most women (Fem), STEM courses, and courses that tend to appeal to most men (Masc). Based on these groups, we consider the rating block model
\begin{equation}
\rateblockmat = \left[
\begin{tabular}{c|ccc}
& Fem & STEM & Masc\\
\midrule
W & 0.8 & 0.2 & 0.2 \\
WS & 0.8 & 0.8 & 0.2 \\
MS & 0.2 & 0.8 & 0.8 \\
M & 0.2 & 0.2 & 0.8
\end{tabular}
\right].
\end{equation}

We also consider two observation block models: one with uniform observation probability across all groups $\takeblockmat^{\textrm{uni}} = [0.4]^{4 \times 3}$ and one with unbalanced observation probability inspired by how students are often encouraged to take certain courses
\begin{equation}
\takeblockmat^{\textrm{bias}} = \left[
\begin{tabular}{c|ccc}
& Fem & STEM & Masc\\
\midrule
W & 0.6 & 0.2 & 0.1 \\
WS & 0.3 & 0.4 & 0.2 \\
MS & 0.1 & 0.3 & 0.5 \\
M & 0.05 & 0.5 & 0.35
\end{tabular}
\right]~.
\end{equation}

We define two different user group distributions: one in which each of the four groups is exactly a quarter of the population, and an imbalanced setting where 0.4 of the population is in W, 0.1 in WS, 0.4 in MS, and 0.1 in M. This heavy imbalance is inspired by some of the severe gender imbalances in certain STEM areas today.

For each experiment, we select an observation matrix and user group distribution, generate 400 users and 300 items, and sample preferences and observations of those preferences from the block models. Training on these ratings, we evaluate on the remaining entries of the rating matrix, comparing the predicted rating against the true expected rating, $2\rateblock_{(g_i, h_j)} - 1$.

\subsubsection{Unfairness from different types of underrepresentation}

\begin{figure}[tbp]
\centering
\includegraphics[width=0.32\textwidth]{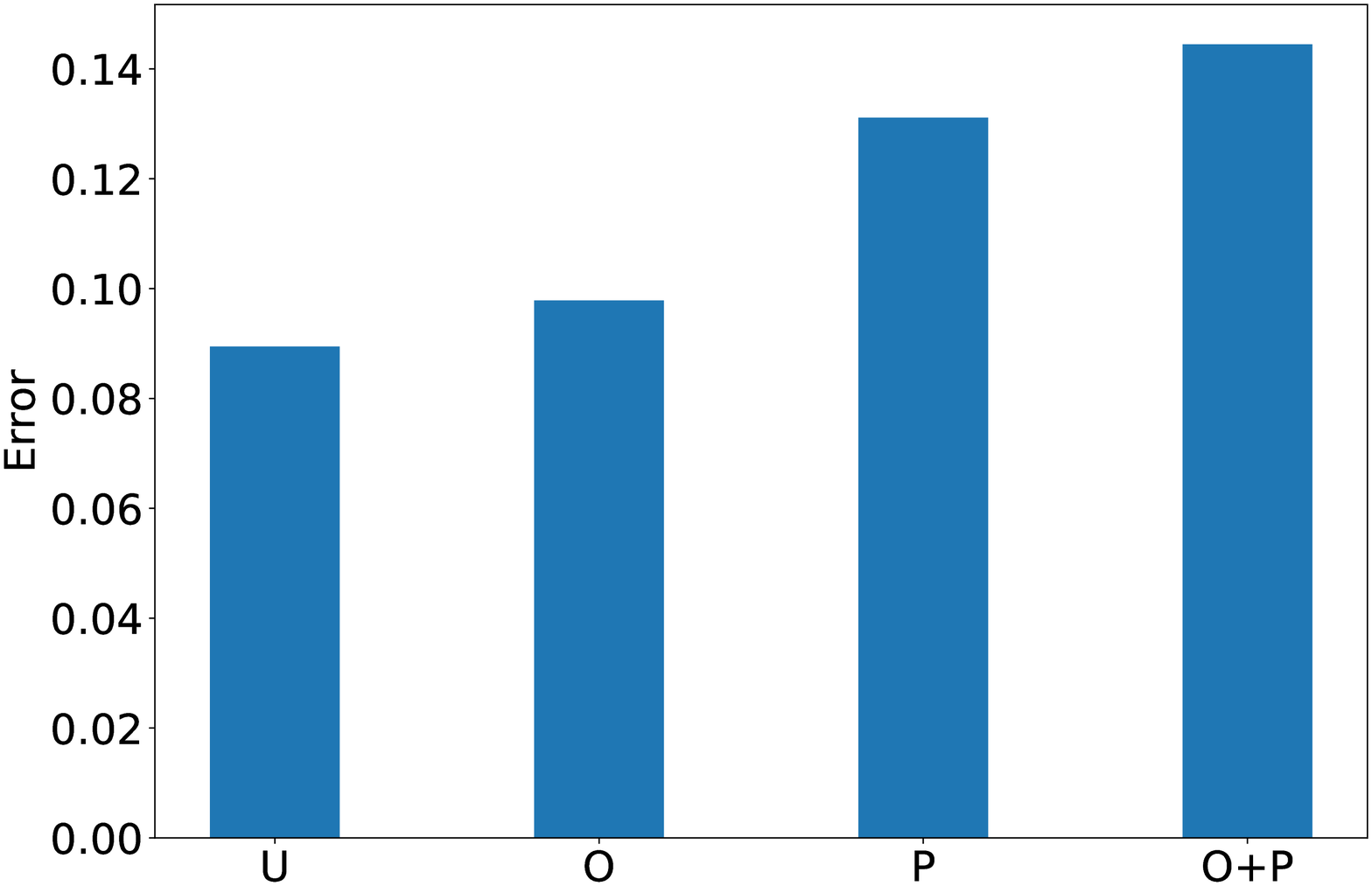}
\includegraphics[width=0.32\textwidth]{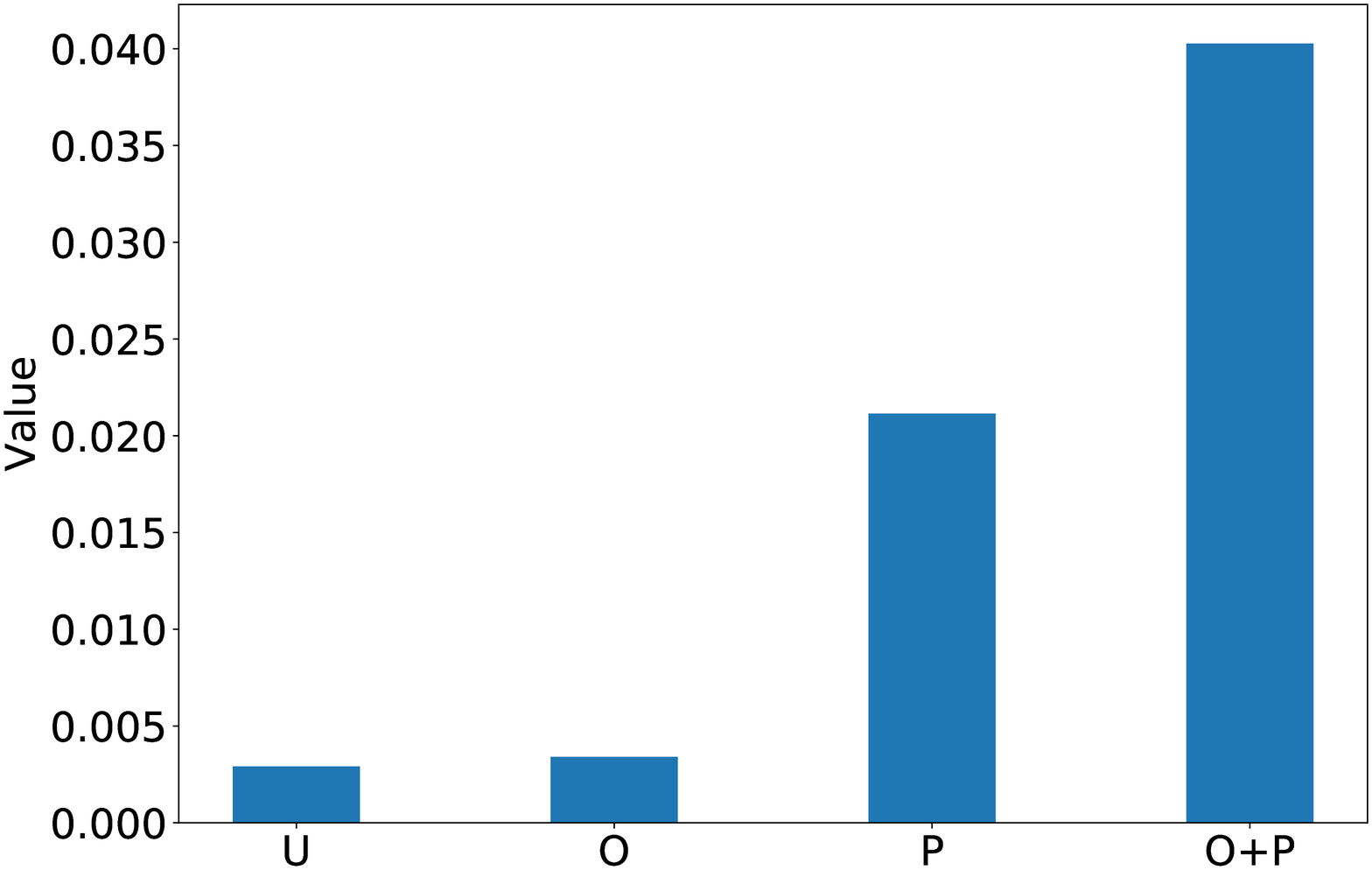}
\includegraphics[width=0.32\textwidth]{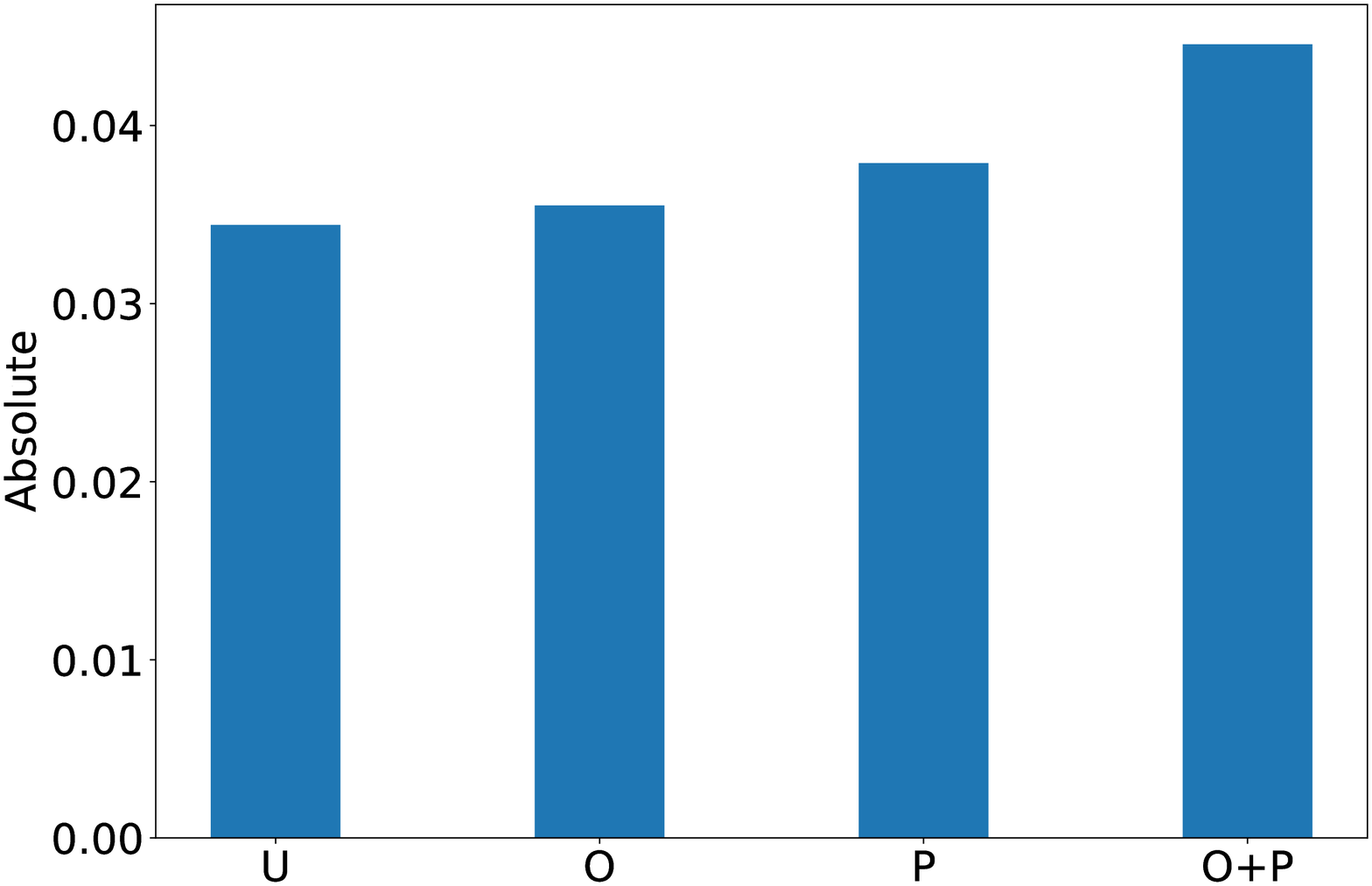}
\includegraphics[width=0.32\textwidth]{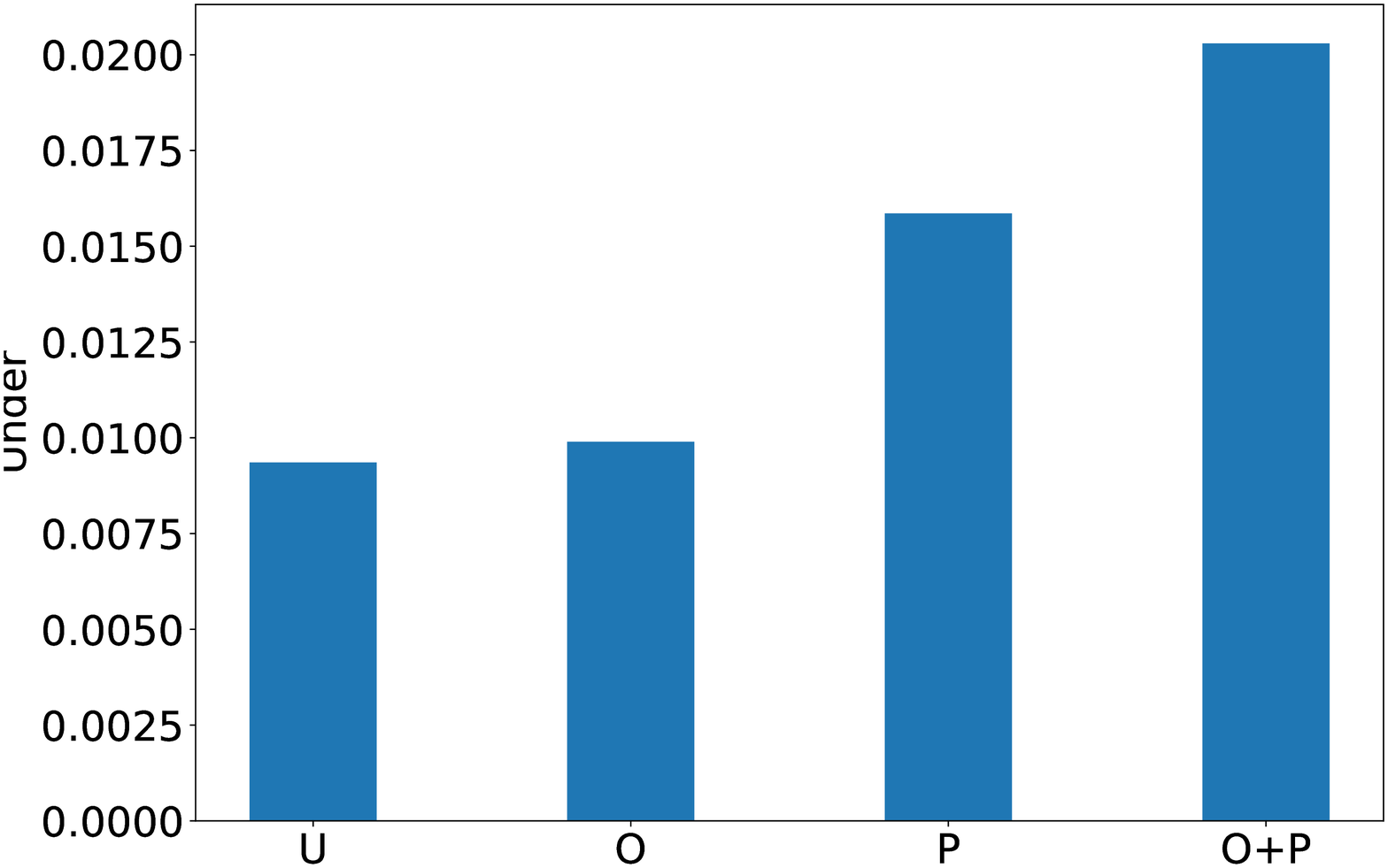}
\includegraphics[width=0.32\textwidth]{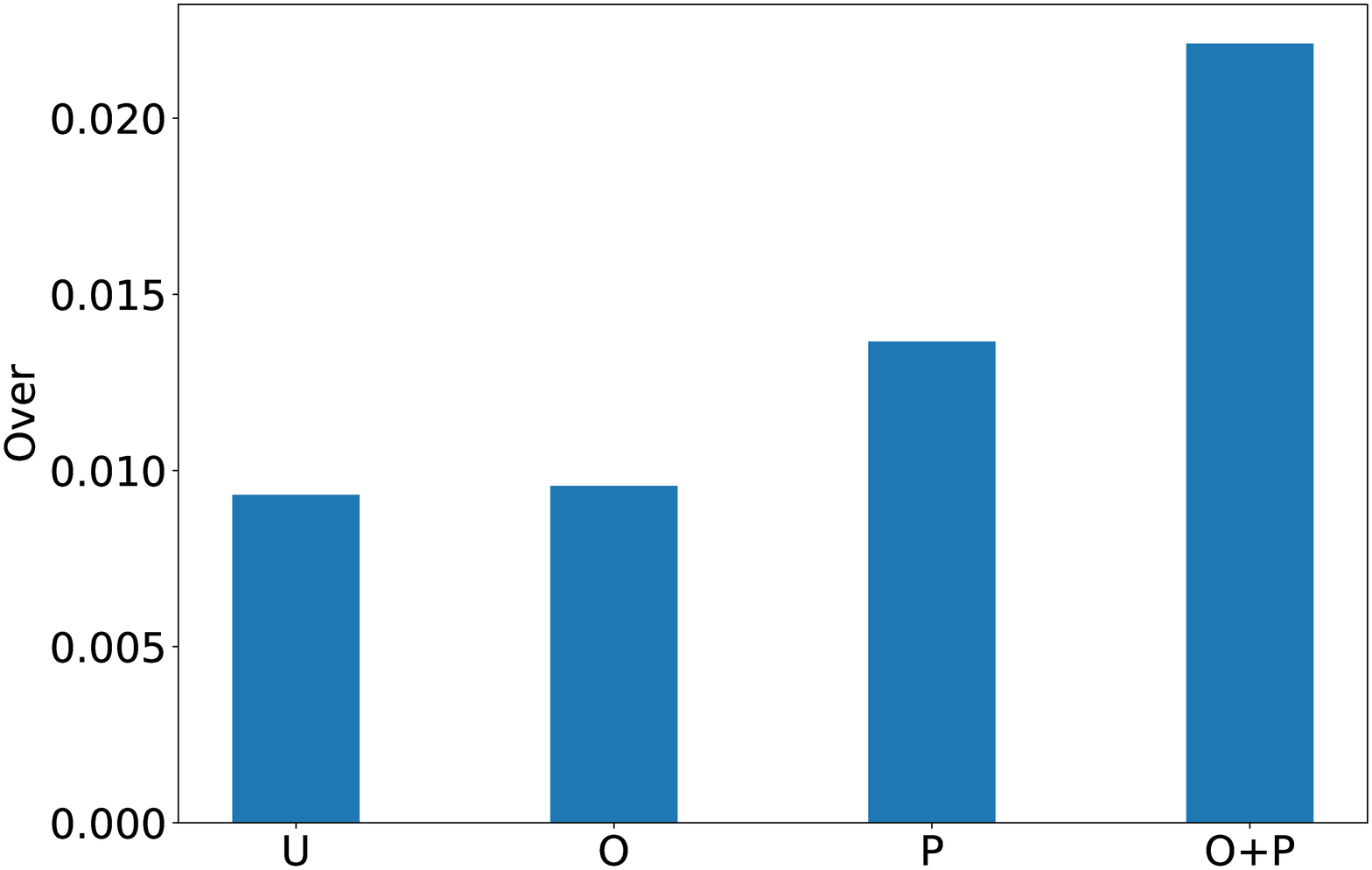}
\includegraphics[width=0.32\textwidth]{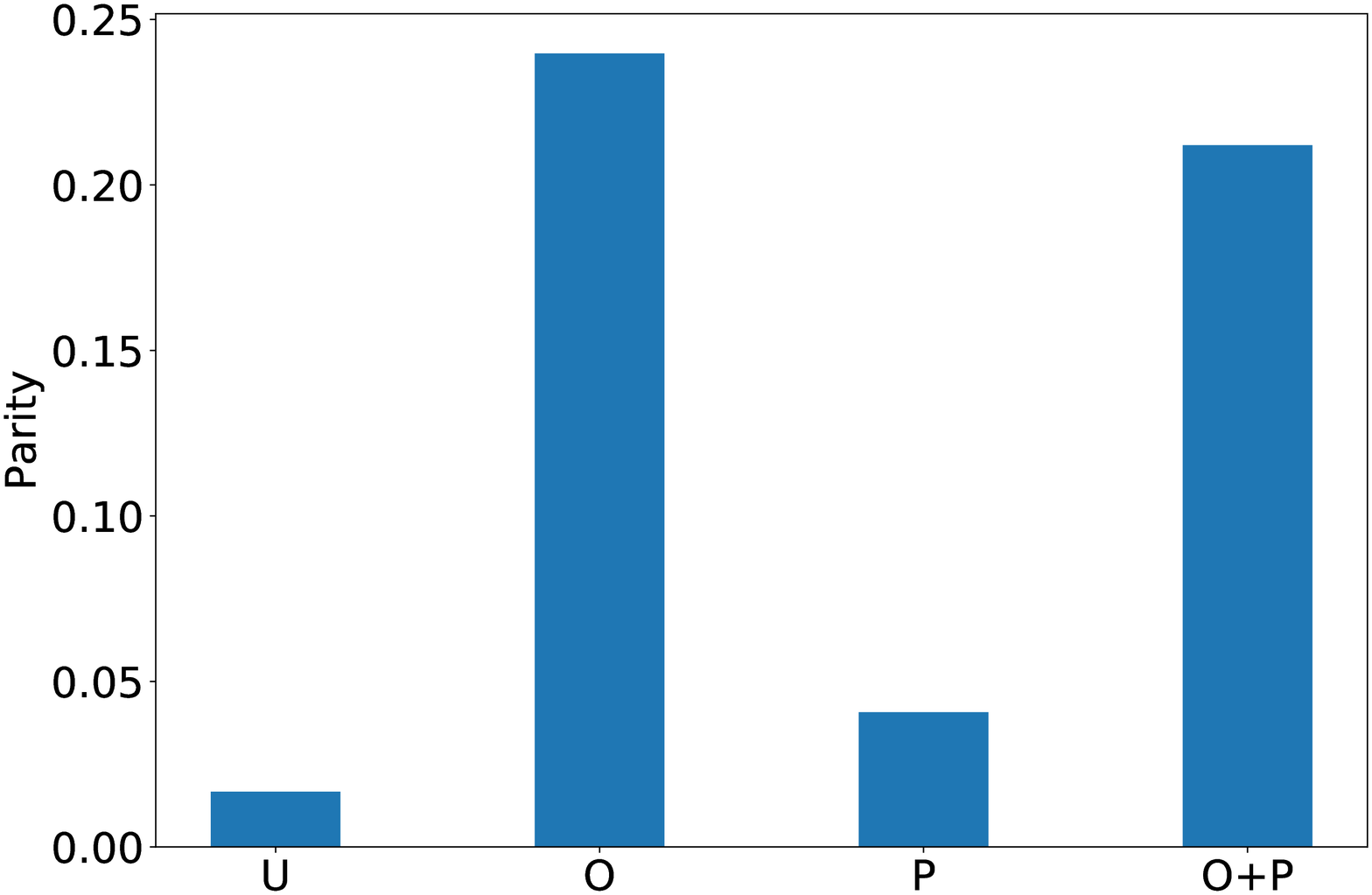}
\caption{Average unfairness scores for standard matrix factorization on synthetic data generated from different underrepresentation schemes. For each metric, the four sampling schemes are uniform (U), biased observations (O), biased populations (P), and both biases (O+P). The reconstruction error and the first four unfairness metrics follow the same trend, while non-parity exhibits different behavior.}
\label{fig:bars}
\end{figure}

Using standard matrix factorization, we measure the various unfairness metrics under the different sampling conditions. We average over five random trials and plot the average score in \cref{fig:bars}. We label the settings as follows: uniform user groups and uniform observation probabilities (U), uniform groups and biased observation probabilities (O), biased user group populations and uniform observations (P), and biased populations and biased observations (P+O).

The statistics demonstrate that each type of underrepresentation contributes to various forms of unfairness. For all metrics except parity, there is a strict order of unfairness: uniform data is the most fair; biased observations is the next most fair; biased populations is worse; and biasing the populations and observations causes the most unfairness. The squared rating error also follows this same trend. In contrast, non-parity behaves differently, in that it is heavily amplified by biased observations but seems unaffected by biased populations. Note that though non-parity is high when the observations are imbalanced, because of the imbalance in the observations, one should actually expect non-parity in the labeled ratings, so it a high non-parity score does not necessarily indicate an unfair situation. The other unfairness metrics, on the other hand, describe examples of unfair behavior by the rating predictor. These tests verify that unfairness can occur with imbalanced populations or observations, even when the measured ratings accurately represent user preferences.

\subsubsection{Optimization of unfairness metrics}

As before, we generate rating data using the block model under the most imbalanced setting: The user populations are imbalanced, and the sampling rate is skewed. We provide the sampled ratings to the matrix factorization algorithms and evaluate on the remaining entries of the expected rating matrix. We again use two-dimensional vectors to represent the users and items, a regularization term of $\lambda = 10^{-3}$, and optimize for 250 iterations using the full gradient. We generate three datasets each and measure squared reconstruction error and the six unfairness metrics.

The results are listed in \cref{tab:synthetic}. For each metric, we print in bold the best average score and any scores that are not statistically significantly distinct according to paired t-tests with threshold 0.05. The results indicate that the learning algorithm successfully minimizes the unfairness penalties, generalizing to unseen, held-out user-item pairs. And reducing any unfairness metric does not lead to a significant increase in reconstruction error. 

The complexity of computing the unfairness metrics is similar to that of the error computation, which is linear in the number of ratings, so adding the fairness term approximately doubles the training time. In our implementation, learning with fairness terms takes longer because loops and backpropagation introduce extra overhead. For example, with synthetic data of $400$ users and $300$ items, it takes $13.46$ seconds to train a matrix factorization model without any unfairness term and $43.71$ seconds for one with value unfairness.

While optimizing each metric leads to improved performance on itself (see the highlighted entries in \cref{tab:synthetic}), a few trends are worth noting. Optimizing any of our new unfairness metrics almost always reduces the other forms of unfairness. An exception is that optimizing absolute unfairness leads to an increase in underestimation. Value unfairness is closely related to underestimation and overestimation, since optimizing value unfairness is even more effective at reducing underestimation and overestimation than directly optimizing them. Also, optimizing value and overestimation are more effective in reducing absolute unfairness than directly optimizing it. Finally, optimizing parity unfairness leads to increases in all unfairness metrics except absolute unfairness and parity itself. These relationships among the metrics suggest a need for practitioners to decide which types of fairness are most important for their applications.


\begin{table}[tbp]
\caption{Average error and unfairness metrics for synthetic data using different fairness objectives. The best scores and those that are statistically indistinguishable from the best are printed in bold. Each row represents a different unfairness penalty, and each column is the measured metric on the expected value of unseen ratings.}
\label{tab:synthetic}
\centering
\vspace{0.1cm}
{\scriptsize
\begin{tabular}{lllllll}
\toprule
Unfairness & Error & Value & Absolute & Underestimation & Overestimation & Non-Parity \\
\midrule
 None &  0.317 $\pm$ 1.3e-02 & 0.649 $\pm$ 1.8e-02 & 0.443 $\pm$ 2.2e-02 & 0.107 $\pm$ 6.5e-03 & 0.544 $\pm$ 2.0e-02 & 0.362 $\pm$ 1.6e-02 \\
 Value &  \textbf{0.130 $\pm$ 1.0e-02} & \hl{\textbf{0.245 $\pm$ 1.4e-02}} & 0.177 $\pm$ 1.5e-02 & \textbf{0.063 $\pm$ 4.1e-03} & \textbf{0.199 $\pm$ 1.5e-02} & 0.324 $\pm$ 1.2e-02 \\
 Absolute &  0.205 $\pm$ 8.8e-03 & 0.535 $\pm$ 1.6e-02 & \hl{0.267 $\pm$ 1.3e-02} & 0.135 $\pm$ 6.2e-03 & 0.400 $\pm$ 1.4e-02 & 0.294 $\pm$ 1.0e-02 \\
 Under &  0.269 $\pm$ 1.6e-02 & 0.512 $\pm$ 2.3e-02 & 0.401 $\pm$ 2.4e-02 & \hl{\textbf{0.060 $\pm$ 3.5e-03}} & 0.456 $\pm$ 2.3e-02 & 0.357 $\pm$ 1.6e-02 \\
 Over &  \textbf{0.130 $\pm$ 6.5e-03} & 0.296 $\pm$ 1.2e-02 & \textbf{0.172 $\pm$ 1.3e-02} & 0.074 $\pm$ 6.0e-03 & \hl{0.228 $\pm$ 1.1e-02} & 0.321 $\pm$ 1.2e-02 \\
 Non-Parity &  0.324 $\pm$ 1.3e-02 & 0.697 $\pm$ 1.8e-02 & 0.453 $\pm$ 2.2e-02 & 0.124 $\pm$ 6.9e-03 & 0.573 $\pm$ 1.9e-02 & \hl{\textbf{0.251 $\pm$ 1.0e-02}} \\
\bottomrule
\end{tabular}
}
\end{table}

\subsection{Real Data}

We use the Movielens Million Dataset \citep{harper2016movielens}, which contains ratings (from 1 to 5) by 6,040 users of 3,883 movies. The users are annotated with demographic variables including gender, and the movies are each annotated with a set of genres. We manually selected genres that feature different forms of gender imbalance and only consider movies that list these genres. Then we filter the users to only consider those who rated at least 50 of the selected movies.

The genres we selected are \emph{action}, \emph{crime}, \emph{musical}, \emph{romance}, and \emph{sci-fi}. We selected these genres because they each have a noticeable gender effect in the data.
Women rate musical and romance films higher and more frequently than men. Women and men both score action, crime, and sci-fi films about equally, but men rate these film much more frequently. \Cref{tab:genres} lists these statistics in detail.
After filtering by genre and rating frequency, we have 2,953 users and 1,006 movies in the dataset.

\begin{table}[tbp]
\caption{Gender-based statistics of movie genres in Movielens data.}
\label{tab:genres}
\centering
{\scriptsize
\begin{tabular}{lrrrrr}
\toprule
& Romance & Action & Sci-Fi & Musical & Crime\\
\midrule
Count & 325 & 425 & 237 & 93 & 142\\
Ratings per female user & 54.79 & 52.00 & 31.19 & 15.04 & 17.45\\
Ratings per male user & 36.97 & 82.97 & 50.46 & 10.83 & 23.90 \\
Average rating by women & 3.64 & 3.45 & 3.42 & 3.79 & 3.65\\
Average rating by men & 3.55 & 3.45 & 3.44 & 3.58 & 3.68\\
\bottomrule
\end{tabular}
}
\end{table}

We run five trials in which we randomly split the ratings into training and testing sets, train each objective function on the training set, and evaluate each metric on the testing set. The average scores are listed in \cref{tab:movielens}, where bold scores again indicate being statistically indistinguishable from the best average score.
On real data, the results show that optimizing each unfairness metric leads to the best performance on that metric without a significant change in the reconstruction error. As in the synthetic data, optimizing value unfairness leads to the most decrease on under- and overestimation. Optimizing non-parity again causes an increase or no change in almost all the other unfairness metrics. 

\begin{table}[tbp]
\caption{Average error and unfairness metrics for movie-rating data using different fairness objectives.}
\label{tab:movielens}
\centering
\vspace{0.1cm}
{\scriptsize
\begin{tabular}{lllllll}
\toprule
Unfairness & Error & Value & Absolute & Underestimation & Overestimation & Non-Parity \\
\midrule
 None &  0.887 $\pm$ 1.9e-03 & 0.234 $\pm$ 6.3e-03 & 0.126 $\pm$ 1.7e-03 & 0.107 $\pm$ 1.6e-03 & 0.153 $\pm$ 3.9e-03 & 0.036 $\pm$ 1.3e-03 \\
 Value &  0.886 $\pm$ 2.2e-03 & \hl{\textbf{0.223 $\pm$ 6.9e-03}} & 0.128 $\pm$ 2.2e-03 & \textbf{0.102 $\pm$ 1.9e-03} & \textbf{0.148 $\pm$ 4.9e-03} & 0.041 $\pm$ 1.6e-03 \\
 Absolute &  0.887 $\pm$ 2.0e-03 & 0.235 $\pm$ 6.2e-03 & \hl{\textbf{0.124 $\pm$ 1.7e-03}} & 0.110 $\pm$ 1.8e-03 & 0.151 $\pm$ 4.2e-03 & 0.023 $\pm$ 2.7e-03 \\
 Under &  0.888 $\pm$ 2.2e-03 & 0.233 $\pm$ 6.8e-03 & 0.128 $\pm$ 1.8e-03 & \hl{\textbf{0.102 $\pm$ 1.7e-03}} & 0.156 $\pm$ 4.2e-03 & 0.058 $\pm$ 9.3e-04 \\
 Over &  \textbf{0.885 $\pm$ 1.9e-03} & 0.234 $\pm$ 5.8e-03 & \textbf{0.125 $\pm$ 1.6e-03} & 0.112 $\pm$ 1.9e-03 & \hl{\textbf{0.148 $\pm$ 4.1e-03}} & 0.015 $\pm$ 2.0e-03 \\
 Non-Parity &  0.887 $\pm$ 1.9e-03 & 0.236 $\pm$ 6.0e-03 & 0.126 $\pm$ 1.6e-03 & 0.110 $\pm$ 1.7e-03 & 0.152 $\pm$ 3.9e-03 & \hl{\textbf{0.010 $\pm$ 1.5e-03}} \\
\bottomrule
\end{tabular}
}
\end{table}

\section{Conclusion}
\label{sec:conclusion}

In this paper, we discussed various types of unfairness that can occur in collaborative filtering. We demonstrate that these forms of unfairness can occur even when the observed rating data is correct, in the sense that it accurately reflects the preferences of the users. We identify two forms of data bias that can lead to such unfairness. We then demonstrate that augmenting matrix-factorization objectives with these unfairness metrics as penalty functions enables a learning algorithm to minimize each of them. Our experiments on synthetic and real data show that minimization of these forms of unfairness is possible with no significant increase in reconstruction error.

We also demonstrate a combined objective that penalizes both overestimation and underestimation. Minimizing this objective leads to small unfairness penalties for the other forms of unfairness. Using this combined objective may be a good approach for practitioners. However, no single objective was the best for all unfairness metrics, so it remains necessary for practitioners to consider precisely which form of fairness is most important in their application and optimize that specific objective.

\paragraph{Future Work}

While our work in this paper focused on improving fairness among users so that the model treats different groups of users fairly, we did not address fair treatment of different item groups. The model could be biased toward certain items, e.g., performing better at prediction for some items than others in terms of accuracy or over- and underestimation. Achieving fairness for both users and items may be important when considering that the items may also suffer from discrimination or bias, for example, when courses are taught by instructors with different demographics. 

Our experiments demonstrate that minimizing empirical unfairness generalizes, but this generalization is dependent on data density. When ratings are especially sparse, the empirical fairness does not always generalize well to held-out predictions. We are investigating methods that are more robust to data sparsity in future work.

Moreover, our fairness metrics assume that users rate items according to their true preferences. This assumption is likely to be violated in real data, since ratings can also be influenced by various environmental factors. E.g., in education, a student's rating for a course also depends on whether the course has an inclusive and welcoming learning environment. However, addressing this type of bias may require additional information or external interventions beyond the provided rating data.

Finally, we are investigating methods to reduce unfairness by directly modeling the two-stage sampling process we used to generate synthetic, biased data. We hypothesize that by explicitly modeling the rating and observation probabilities as separate variables, we may be able to derive a principled, probabilistic approach to address these forms of data imbalance.

\bibliographystyle{abbrv}
\bibliography{yao-nips17}

\begin{thebibliography}{10}

\bibitem{beede2011women}
D.~N. Beede, T.~A. Julian, D.~Langdon, G.~McKittrick, B.~Khan, and M.~E. Doms.
\newblock Women in {STEM}: A gender gap to innovation.
\newblock {\em U.S. Department of Commerce, Economics and Statistics
  Administration}, 2011.

\bibitem{beutel2017beyond}
A.~Beutel, E.~H. Chi, Z.~Cheng, H.~Pham, and J.~Anderson.
\newblock Beyond globally optimal: Focused learning for improved
  recommendations.
\newblock In {\em Proceedings of the 26th International Conference on World
  Wide Web}, pages 203--212. International World Wide Web Conferences Steering
  Committee, 2017.

\bibitem{broad2014recruiting}
S.~Broad and M.~McGee.
\newblock Recruiting women into computer science and information systems.
\newblock {\em Proceedings of the Association Supporting Computer Users in
  Education Annual Conference}, pages 29--40, 2014.

\bibitem{chausson2010watches}
O.~Chausson.
\newblock Who watches what? {A}ssessing the impact of gender and personality on
  film preferences.
\newblock {\em
  http://mypersonality.org/wiki/doku.php?id=movie\_tastes\_and\_personality},
  2010.

\bibitem{dascalu2016educational}
M.-I. Dascalu, C.-N. Bodea, M.~N. Mihailescu, E.~A. Tanase, and
  P.~Ordo{\~n}ez~de Pablos.
\newblock Educational recommender systems and their application in lifelong
  learning.
\newblock {\em Behaviour \& Information Technology}, 35(4):290--297, 2016.

\bibitem{daymont1984job}
T.~N. Daymont and P.~J. Andrisani.
\newblock Job preferences, college major, and the gender gap in earnings.
\newblock {\em Journal of Human Resources}, pages 408--428, 1984.

\bibitem{ekstrand2011collaborative}
M.~D. Ekstrand, J.~T. Riedl, J.~A. Konstan, et~al.
\newblock Collaborative filtering recommender systems.
\newblock {\em Foundations and Trends in Human-Computer Interaction},
  4(2):81--173, 2011.

\bibitem{griffith2010persistence}
A.~L. Griffith.
\newblock Persistence of women and minorities in {STEM} field majors: Is it the
  school that matters?
\newblock {\em Economics of Education Review}, 29(6):911--922, 2010.

\bibitem{hardt2016equality}
M.~Hardt, E.~Price, N.~Srebro, et~al.
\newblock Equality of opportunity in supervised learning.
\newblock In {\em Advances in Neural Information Processing Systems}, pages
  3315--3323, 2016.

\bibitem{harper2016movielens}
F.~M. Harper and J.~A. Konstan.
\newblock The {M}ovielens datasets: History and context.
\newblock {\em ACM Transactions on Interactive Intelligent Systems (TiiS)},
  5(4):19, 2016.

\bibitem{holland1976local}
P.~W. Holland and S.~Leinhardt.
\newblock Local structure in social networks.
\newblock {\em Sociological Methodology}, 7:1--45, 1976.

\bibitem{huber1964robust}
P.~J. Huber.
\newblock Robust estimation of a location parameter.
\newblock {\em The Annals of Mathematical Statistics}, pages 73--101, 1964.

\bibitem{kamishima2012enhancement}
T.~Kamishima, S.~Akaho, H.~Asoh, and J.~Sakuma.
\newblock Enhancement of the neutrality in recommendation.
\newblock In {\em Decisions@ RecSys}, pages 8--14, 2012.

\bibitem{kamishima2013efficiency}
T.~Kamishima, S.~Akaho, H.~Asoh, and J.~Sakuma.
\newblock Efficiency improvement of neutrality-enhanced recommendation.
\newblock In {\em Decisions@ RecSys}, pages 1--8, 2013.

\bibitem{kamishima2014correcting}
T.~Kamishima, S.~Akaho, H.~Asoh, and J.~Sakuma.
\newblock Correcting popularity bias by enhancing recommendation neutrality.
\newblock In {\em RecSys Posters}, 2014.

\bibitem{kamishima2016model}
T.~Kamishima, S.~Akaho, H.~Asoh, and I.~Sato.
\newblock Model-based approaches for independence-enhanced recommendation.
\newblock In {\em Data Mining Workshops (ICDMW), 2016 IEEE 16th International
  Conference on}, pages 860--867. IEEE, 2016.

\bibitem{kamishima2011fairness}
T.~Kamishima, S.~Akaho, and J.~Sakuma.
\newblock Fairness-aware learning through regularization approach.
\newblock In {\em 11th International Conference on Data Mining Workshops
  (ICDMW)}, pages 643--650. IEEE, 2011.

\bibitem{kingma2014adam}
D.~Kingma and J.~Ba.
\newblock Adam: A method for stochastic optimization.
\newblock {\em arXiv preprint arXiv:1412.6980}, 2014.

\bibitem{koren2009matrix}
Y.~Koren, R.~Bell, and C.~Volinsky.
\newblock Matrix factorization techniques for recommender systems.
\newblock {\em Computer}, 42(8), 2009.

\bibitem{lum2016statistical}
K.~Lum and J.~Johndrow.
\newblock A statistical framework for fair predictive algorithms.
\newblock {\em arXiv preprint arXiv:1610.08077}, 2016.

\bibitem{marlin2012collaborative}
B.~Marlin, R.~S. Zemel, S.~Roweis, and M.~Slaney.
\newblock Collaborative filtering and the missing at random assumption.
\newblock {\em arXiv preprint arXiv:1206.5267}, 2012.

\bibitem{marlin2009collaborative}
B.~M. Marlin and R.~S. Zemel.
\newblock Collaborative prediction and ranking with non-random missing data.
\newblock In {\em Proceedings of the third ACM conference on Recommender
  systems}, pages 5--12. ACM, 2009.

\bibitem{pedreshi2008discrimination}
D.~Pedreshi, S.~Ruggieri, and F.~Turini.
\newblock Discrimination-aware data mining.
\newblock In {\em Proceedings of the 14th ACM SIGKDD International Conference
  on Knowledge Discovery and Data Mining}, pages 560--568. ACM, 2008.

\bibitem{sacin2009recommendation}
C.~V. Sacin, J.~B. Agapito, L.~Shafti, and A.~Ortigosa.
\newblock Recommendation in higher education using data mining techniques.
\newblock In {\em Educational Data Mining}, 2009.

\bibitem{sahebi2015takes}
S.~Sahebi and P.~Brusilovsky.
\newblock It takes two to tango: An exploration of domain pairs for
  cross-domain collaborative filtering.
\newblock In {\em Proceedings of the 9th ACM Conference on Recommender
  Systems}, pages 131--138. ACM, 2015.

\bibitem{smith2011women}
E.~Smith.
\newblock Women into science and engineering? {G}endered participation in
  higher education {STEM} subjects.
\newblock {\em British Educational Research Journal}, 37(6):993--1014, 2011.

\bibitem{thai2010recommender}
N.~Thai-Nghe, L.~Drumond, A.~Krohn-Grimberghe, and L.~Schmidt-Thieme.
\newblock Recommender system for predicting student performance.
\newblock {\em Procedia Computer Science}, 1(2):2811--2819, 2010.

\bibitem{zafar2017fairness}
M.~B. Zafar, I.~Valera, M.~Gomez~Rodriguez, and K.~P. Gummadi.
\newblock Fairness constraints: Mechanisms for fair classification.
\newblock {\em arXiv preprint arXiv:1507.05259}, 2017.

\bibitem{zemel2013learning}
R.~Zemel, Y.~Wu, K.~Swersky, T.~Pitassi, and C.~Dwork.
\newblock Learning fair representations.
\newblock In {\em Proceedings of the 30th International Conference on Machine
  Learning}, pages 325--333, 2013.

\end{thebibliography}

\end{document}